\documentclass[11pt]{article}   
\usepackage{graphicx}				
\usepackage{cancel}
\usepackage{amsthm}
\usepackage{tikz}
\usepackage{pgfplots}
\usetikzlibrary{decorations.markings}

\usepackage{enumitem}
\usepackage{xcolor}
\definecolor{mydarkgreen}{RGB}{0,100,0}

\renewcommand{\i}{\mathrm{i}}

\usepackage{hyperref}
\usepackage{amssymb}
\usepackage{amsmath}
\usepackage{amsfonts}
\usepackage{mathbbol}

\usepackage{color}

\newcommand{\blue}{\color{black}}
\newcommand{\red}{\color{black}}

\newcommand{\redd}{\color{black}}
\newcommand{\bluenew}{\color{black}}
\newcommand{\blues}{\color{black}}

\newlength{\dinwidth}
\newlength{\dinmargin}

 \usepackage{cancel}

\newcommand{\one}{\mathbb{1}}
\newcommand{\mrm}{\mathrm}

\newcommand{\ka}{\kappa}

\newcommand{\wh}{\widehat}

\newcommand{\Si}{\Sigma}

\newcommand{\De}{\Delta}

\newcommand{\La}{\Lambda}

\newcommand{\nin}{\noindent}
\newcommand{\lan}{\langle}
\newcommand{\ran}{\rangle}

\newcommand{\ti}{\tilde}

\newcommand{\pa}{\partial}

\newcommand{\h}{\fr{1}{2}}
\newcommand{\e}{\mathrm{e}}

\newcommand{\vp}{\varphi}

\newcommand{\de}{\delta}

\newcommand{\nat}{\mathbb{N}}

\newcommand{\eps}{\varepsilon}
\newcommand{\fr}[2]{\frac{#1}{#2}}
\newcommand{\al}{\alpha}
\newcommand{\be}{\beta}

\newcommand{\real}{\mathbb{R}}

\newcommand{\la}{\lambda}

\newcommand{\non}{\nonumber}

\def\proof{\noindent{\bf Proof. }}
\def\qed{$\Box$\medskip}

\newtheorem{theoreme}{Theorem}[section]
\newtheorem{proposition}[theoreme]{Proposition}
\newtheorem{lemma}[theoreme]{Lemma}
\newtheorem{definition}[theoreme]{Definition}
\newtheorem{corollary}[theoreme]{Corollary}
\newtheorem{remark}[theoreme]{Remark}
\newtheorem{example}[theoreme]{Example}
\newtheorem{criterion}[theoreme]{Criterion}
\newtheorem{assumption}[theoreme]{Assumption}
\newcommand{\bem}{\begin{multline}}
\newcommand{\eem}{\end{multline}}

\newcommand{\beex}{\begin{example}}
\newcommand{\eeex}{\end{example}}
\newcommand{\bea}{\begin{assumption}}
\newcommand{\eea}{\end{assumption}}
\newcommand{\beq}{\begin{equation}}
\newcommand{\eeq}{\end{equation}}
\newcommand\beqa{\begin{eqnarray}}   
\newcommand\eeqa{\end{eqnarray}} 
\newcommand{\ben}{\begin{arabicenumerate}}
\newcommand{\een}{\end{arabicenumerate}}
\newcommand{\bex}{\begin{example}}
\newcommand{\eex}{\end{example}}
\newcommand{\ber}{\begin{remark}}
\newcommand{\eer}{\end{remark}}
\newcommand{\bec}{\begin{corollary}}
\newcommand{\eec}{\end{corollary}}
\newcommand{\bed}{\begin{definition}}
\newcommand{\eed}{\end{definition}}
\newcommand{\bep}{\begin{proposition}}
\newcommand{\eep}{\end{proposition}}
\newcommand{\becr}{\begin{criterion}}
\newcommand{\eecr}{\end{criterion}}
\def\bel{\begin{lemma}}
\def\eel{\end{lemma}}
\def\bet{\begin{theoreme}}
\def\eet{\end{theoreme}}
\def\bed{\begin{definition}}
\def\eed{\end{definition}}

\newcommand{\N}{\La}

\newcommand{\2}{\!\!\!&}

\renewcommand{\i}{\mathrm{i}}
\newcommand{\V}{\mathcal{V}}
\renewcommand{\S}{B}
\newcommand{\tiC}{\tilde{C}}					
\newcommand{\thetas}{\F}

\newcommand{\Sm}{\Si}
\newcommand{\F}{V}

\title{Exact Schwinger functions for a class of bounded interactions in $d\geq 2$}

\author{Wojciech Dybalski   \\[5mm]
\normalsize  Faculty of Mathematics and Computer Science, \\
\normalsize Adam Mickiewicz University in Poznań, \\
\normalsize ul.~Uniwersytetu Pozna\'nskiego 4, 61-614 Pozna\'n, Poland\\
\normalsize E-mail: {\tt wojciech.dybalski@amu.edu.pl} \\ [2mm]
}

\date{}

\begin{document}

\maketitle

\begin{abstract}
We consider a scalar Euclidean QFT with  interaction given by a bounded, measurable
function $\F$ such that  $V^{\pm}:=\lim_{w\to \pm\infty}V(w)$  exist. 
We find a field renormalization such that  all the $n$-point connected Schwinger functions for $n\neq 2$ exist
non-perturbatively in the UV limit.   They coincide with the tree-level 
{\redd one-particle irreducible} Schwinger functions of  the $\mathrm{erf}(\phi/\sqrt{2})$ interaction with a coupling
constant $\fr{1}{2} (V^+ - V^-)$.  By a slight modification  of our construction we can change this coupling constant 
to $\fr{1}{2} (V_+ - V_-)$,  where $V_{\pm}:= \lim_{w\to 0^{\pm}} V(w)$. Thereby, non-Gaussianity of these latter theories is governed by 
 a discontinuity   of $\F$ at zero. The open problem of controlling also the two-point function of these QFTs is discussed.
\end{abstract}

\section{Introduction}
\setcounter{equation}{0}

Rigorous construction of interacting scalar quantum field theories in spacetime dimension $d=2, 3$
is among the most important achievements of the constructive programme \cite{GJ}.  These efforts focussed
on polynomial interactions and were limited by the requirement of super-renormalizability:
For $d=2$ all  (bounded from below) polynomials are covered, for $d=3$ only polynomials up to the
fourth order,  whereas for $d\geq 4$ all the polynomial theories are expected to be trivial \cite{Ai82,  Fr82, AD21, WK25}. These limitations
suggest some room for non-polynomial scalar theories  in $d=2$. In fact,  theories
based on exponential interaction functions, such as the sine-Gordon and sinh-Gordon model, are
among the classical topics of constructive QFT \cite{FP77} and still constitute a very active field of research \cite{Le08,CT16, HS16,BFR21,BD21,BH22,Ko23,BDR24, CFM24, GM24}. On the other hand, for $d>2$, there seems to be no
room for non-polynomial interactions, at least if we think about them in terms of their
Taylor expansions.  But this leaves aside bounded interaction functions like $\mrm{arctan}(\phi)$,
whose Taylor expansion has finite radius of convergence and thus does not capture the large-field behaviour.
Not to speak of functions like $\mrm{sgn}(\phi)$ which lack a Taylor expansion around zero altogether. Clearly,
polynomial interactions are  not a reliable guide to study such bounded interactions and we are not aware of 
any systematic treatment.    In this short note we
demonstrate that these models are exactly solvable for any $d\geq 2$ in a sense which we now explain. 
 
We  consider interaction functions  $\F: \real\to \real$ which are bounded and measurable. Depending on a situation 
we will need the existence of the following limits:
\begin{align}
\F_{\pm}&:= \lim_{w\to 0^{\pm}} \F(w),     \label{limits-assumption-zero} \tag{$\mrm{A}_{{1}}$} \\
\F^{\pm}&:=\lim_{w\to \pm\infty} \F(w).  \label{limits-assumption}  \tag{$\mrm{A}_{{2}}$}
\end{align}
We will refer to these requirements as assumptions (\ref{limits-assumption-zero}) and (\ref{limits-assumption}), respectively.
The corresponding interaction term in the finite volume $\S:=\big[-\fr{L}{2}, \fr{L}{2} \big]^d$, $L>0$,  is $\V(\phi):=\int_{\S}dx\, \F(\phi(x))$. 
We do not impose any Wick ordering on the interaction.

Next, we introduce the free covariance with mass $m>0$ and a UV cut-off $\N$ s.t. $\log(\La)\geq 1$:	 
\beqa
C_\N(x):=\int_{\real^d} \fr{dp}{\red (2\pi)^d}\,\e^{\i px} \, \fr{\e^{-(p^2+m^2)/\N^2}}{p^2+m^2}, \quad C(x):=\lim_{\N\to \infty}C_\N(x). \label{covariance}
\eeqa
We allow for a field renormalization $Z_\N>0$ and denote the renormalized covariance by $\tiC_\N:=Z_\N C_\N$.
Let $\nu_{\tiC_\N}$ be the corresponding Gaussian measure with mean zero and denote the expectation w.r.t. this measure 
by $\lan \ldots \ran_{\tiC_\N}=\int \ldots d\nu_{\tiC_\N}$.
In particular,
\beqa
S_{0,\N}(J):= \lan \e^{\phi(J)}\ran_{\tiC_\N}=\e^{\h \lan J, \tiC_\N J\ran},\quad J\in S(\real^d;\real) \label{Fourier-transform-measure}
\eeqa
is the generating functional of the Schwinger functions of the free theory, where   $\lan \cdot, \cdot\ran$ on the r.h.s. is the scalar product in $L^2(\real^d;\real)$. The generating functional of the interacting theory has the form
\beqa
S_{\N}(J):=\lan \e^{\red \phi(J)} \e^{-\la \V(\phi) } \ran_{\ti{C}_\N}, \quad \la\in \real, \label{perturbed-generating-functional}
\eeqa
{\redd disregarding a $J$-independent normalization constant, which has no effect on connected Schwinger functions}. 
We refer to  Appendix~\ref{Path-integral} for routine arguments {\bluenew showing} the well-definiteness of this functional.
A standard tool to study the effect of the interaction is the following modified generating functional, 
which we state together with its connected counterpart,  cf.  e.g. \cite[Eq. (2.104)]{Sa},
\beqa
\Sm_\N(J):=\fr{S_{\N}(J)}{S_{0,\N}(J)}, \quad   \Sm^{\mrm{c}}_\N(J):=\log \Sm_\N(J).  \label{generating-functionals}
\eeqa
It is our main result, stated in Theorem~\ref{main-theorem} {\bluenew below},  that $\Sm^{\mrm{c}}_\N(J)$ has a finite  and exactly computable  UV limit.   A proof is given in Section~\ref{analysis}. 
\bet\label{main-theorem} Set $Z_\N=C_\N(0)^{\eta}$, $\eta\in \real$. Then, the limit $\Sm^{\mrm{c}}(J):=\lim_{\N\to \infty}\Sm^{\mrm{c}}_\N(J)$ exists for any $J\in S(\real^d;\real)$  under the assumptions  specified below. It has the form:
\begin{enumerate}[label={(\alph*)},itemindent=1em]

\item \label{eta-large-negative} Let $-\infty <\eta<-1$. Then, assuming (\ref{limits-assumption-zero}), 
\beqa
\Sm^{\mrm{c}}(J)={\red-} \la |\S|\fr{\F_++\F_-}{2}.
\eeqa

\item \label{eta-minus-one} Let $\eta=-1$. Then, without additional assumptions, 
\beqa
\Sm^{\mrm{c}}(J)=-\la |\S| \fr{1}{(2\pi)^{1/2}} \int_{\real} dw \, \F(w) \e^{-\h w^2}.  
\eeqa

\item \label{eta-middle} Let $-1<\eta<1$. Then, assuming (\ref{limits-assumption}), 
\beqa
\Sm^{\mrm{c}}(J)={\red-} \la  |\S|\fr{\F^++\F^-}{2}.
\eeqa

\item\label{eta-one} Let $\eta=1$. Then, assuming (\ref{limits-assumption}),
\beqa
\Sm^{\mrm{c}}(J)={\red-} \la |\S|\fr{\F^++\F^-}{2}  {\red -}\la \fr{\F^+-\F^-}{2}   \fr{1}{(2\pi)^{1/2}}\int_{\S} dx\,\int dw\, \mrm{sgn}(w) \e^{-\h (w- \lan \de_x,  C  J\ran )^2}.
\label{Generating-functional-thm}
\eeqa

\item\label{eta-large-positive} Let $1<\eta<\infty$. Then, assuming  (\ref{limits-assumption}), and choosing $J$ compactly supported in Fourier space,
\beqa
\Sm^{\mrm{c}}(J)=-\la |\S| \fr{\F^++\F^-}{2}   
   -\la  \fr{\F^+-\F^-}{2} \one( J \neq 0)\int_{\S} dx\,  \mrm{sgn}(  \lan  \de_{x}, CJ\ran).
\eeqa

\end{enumerate}

\eet
{\blues Since we wanted to steer away from the  trivialization theorems  \cite{Ai82,  Fr82, AD21, WK25}, the philosophy behind Theorem~\ref{main-theorem}  differs from the usual, Wilsonian renormalization.
The field renormalization constants $Z_{\La}$ are not computed from intricate relations linking the  
degrees of freedom at different scales, cf. \cite[Section~4]{FFS92}. Instead, we start from all possible functions  $\La\mapsto Z_{\La}$ 
and try to select those, which give  interesting UV limits of the generating functional (\ref{generating-functionals}). This strategy is reminiscent of the
Buchholz-Verch  renormalization, which is a framework for analyzing the short distance structure of any given QFT \cite{BV95}.  
This similarity is particularly visible from the exposition in \cite[Section 2]{Bu96}, where the conventional language of fields and  their $n$-point
functions is used.  However, in the Buchholz-Verch approach the $n$-point functions of a local, relativistic QFT are given and the goal is to compute their scaling limits. Our goal, instead, is to compute such $n$-point functions.}

We note that $\Sm^{\mrm{c}}_\N(J)$ is the generating functional of connected $n$-point Schwinger functions  
$S^{\mrm{c}}_{\N,n}$ of the theory with cut-offs for $n\neq 2$, {\redd cf. \cite[Chapter 3]{PT11}}.  Therefore, we interpret $\Sm^{\mrm{c}}(J)$ as the 
generating functional of the connected Schwinger functions $S^{\mrm{c}}_{n\neq 2}$ of the limiting theory. Part \ref{eta-one}
of  Theorem~\ref{main-theorem} is the most interesting one 
as the  connected Schwinger functions $S^{\mrm{c}}_{n\neq 2}$ of the limiting theory are non-trivial and readily computable,
cf.~Appendix~\ref{Schwinger-functions-app}, 
\beqa
S^{\mrm{c}}_{n}(x_1,\ldots, x_n)\2:=\2 \fr{\de}{\de J(x_1)}  \ldots \fr{\de}{\de J(x_n)}\Sm^{\mrm{c}}(J) |_{J=0}\non\\
\2=\2{\red -}\la \fr{(\F^+-\F^-)}{2}  [(\pa_w)^{n} \mrm{erf}(w/\sqrt{2}) ]_{w=0}   \int_{\S}dx \,   C(x-x_1)\ldots C(x-x_n).
\label{Schwinger-functions}
\eeqa
We observe that these $n$-point  functions coincide with the tree-level {\redd one-particle irreducible} Schwinger functions of a perturbative theory with the interaction function $\mrm{erf}(\phi /\sqrt{2})$
and coupling constant $\la \fr{(\F^+-\F^-)}{2}$.  We find it quite surprising that for all the bounded interaction functions $\F$, as defined above, we obtain the same Schwinger functions for $n\neq 2$, up to a finite coupling constant renormalization. 
We also note  that the infinite volume limit $\S\nearrow \real^d$ can easily be taken in (\ref{Schwinger-functions})  after smearing with test functions.
{\blues However, a subsequent $m\to 0$ limit of the one-point function would diverge, so our result does not immediately generalize to perturbations
of  massless free fields.}

 The remaining parts of Theorem~\ref{main-theorem} also carry physically relevant information, although the Schwinger functions $S^{\mrm{c}}_{n\neq 2}$ are zero or undefined. From the point of view of statistical physics,  {\redd $\Sm^{\mrm{c}}(J)$} is the change of the free energy due to interaction, which is an interesting quantity.
Depending on the parameter $\eta$,  it probes various properties  of the interaction function $\F$, such as the behaviour at zero or infinity.    For example,  in the cases \ref{eta-middle}-\ref{eta-large-positive},  we have 
\beqa
\Sm^{\mrm{c}}(0)={\red -} \la  |\S|\fr{(\F^++\F^-)}{2}. \label{UV-stability}
\eeqa
We set $J=0$ here to make this quantity  insensitive to the denominator $S_{0,\N}(J=0)=1$ in (\ref{generating-functionals}) and
thus consistent with the standard definition of the free energy in constructive QFT.  We recall that upper and lower bounds on 
free energy, called UV stability, are difficult to prove  for polynomial interactions, see e.g. \cite{Di13}.  For bounded interactions treated
in this note {\bluenew $\Sm^{\mrm{c}}(0)$} is readily computable.


{\redd  Our paper is organized as follows: Section~\ref{analysis} starts with a non-technical  explanation of a factorization mechanism which is behind Theorem~\ref{main-theorem} and then moves on to the proof of {\bluenew this} result. In Section~\ref{generalisation} we generalize our construction to 
a certain class of $\N$-dependent interaction functions. In Section~\ref{conclusion} we discuss the problem of controlling also  the two-point
Schwinger function for our class of bounded interactions.}

\vspace{0.5cm}

\nin \textbf{Acknowledgements:} I would like to thank Paweł Duch and Yoh Tanimoto for useful comments. Financial  support  
 of the National Science Centre, Poland, via  the grant `Sonata Bis' 2019/34/E/ST1/00053,  is gratefully acknowledged.

\vspace{0.5cm}

\nin\textbf{Data availability:} We do not analyse or generate any datasets, because our work proceeds within a  mathematical approach. One can obtain the relevant materials from the references below.

\vspace{0.5cm}

\nin\textbf{Competing interests:} This research was funded by the National Science Centre, Poland, via  the grant `Sonata Bis' 2019/34/E/ST1/00053.

\section{Analysis} \label{analysis}
\setcounter{equation}{0}

 All constants $c,c_1,c_2,\ldots$ in the following depend only on $m$ and $d$ and may change from line to line. Our proof of Theorem~\ref{main-theorem} is based on the following proposition:
\bep\label{factorization-two}  The following bound holds true for $\ell\in \nat$
\beqa
\2 \2\bigg|\, \fr{\lan  {\blue \e^{\red \phi(J)}} \V(\phi)^{\ell} \ran_{ \ti{C}_\N}}{ {\red  \lan  {\red \e^{\phi(J)} }  \ran_{ \ti{C}_\N}      }  }    - 
 \bigg( \fr{\lan   {\blue \e^{\red \phi(J)} }\V(\phi) \ran_{ \ti{C}_\N}}{\lan  {\red \e^{\phi(J)} }  \ran_{ \ti{C}_\N}  } \bigg)^{\ell}
     \bigg| \leq c \ell^6 \|\F\|^{\ell}_{\infty} |\S|^{\ell} \fr{1}{ \sqrt{\log(\N)}}. \label{factorization-property}
\eeqa
\eep
Let us first explain in a simple example the mechanism, which is behind the factorization property (\ref{factorization-property}).
We choose $\F(w)=\mrm{sgn}(w)$, which is  invariant under scaling $w\mapsto \al w$, $\al>0$. Hence, the Fourier transform
satisfies $\hat{\F}(\al {\redd u})=\al^{-1}\hat{\F}({\redd u})$. Using this property,
 representation $\F(\phi(x))=\fr{1}{\sqrt{2\pi}} \int du\, \hat{\F}(u)\e^{\i \phi(x) u}$ and relation~(\ref{Fourier-transform-measure}), we obtain
\newcommand{\uu}{{\redd u}}
\beqa
\fr{\lan \e^{\red \phi(J)} \V(\phi)^{2} \ran_{ \ti{C}_\N}}{ \lan  {\red \e^{\phi(J)} }  \ran_{ \ti{C}_\N} } =\fr{1}{2\pi}\int_{\S^2} dx_1dx_2 \int_{\real^{2}}d\uu_1 d\uu_2 \, \hat{\F}(\uu_1) \hat{\F}(\uu_2)  \e^{ -\h \sum_{i=1}^2 \uu_i^2- \uu_1 \uu_2 \fr{C_\N(x_1-x_2)}{C_\N(0)}+\i
\fr{\lan \sum_{j=1}^{2}\uu_j\de_{x_j}, \ti{C}_\N J\ran }{\ti{C}_\N(0)^{1/2} }   }. \label{simple-example}
\eeqa
Clearly, the only obstacle to  factorization of the  $u$-integral is the term involving $\fr{C_\N(x_1-x_2)}{C_\N(0)}$.  But for $x_1\neq x_2$
this expression tends to zero with $\N\to \infty$ by Lemma \ref{propagator-lemma-universal} below. On the other hand, the region close to
the diagonal $x_1=x_2$ gives a small contribution for bounded interactions, cf. estimates (\ref{diagonal-one}), (\ref{diagonal-two}) below. 
{\blues (Unfortunately, for unbounded interactions, such as the $P(\phi)_2$ models, these estimates fail.)}
We stress that this factorization mechanism is independent of the field renormalization and relevant to all  parts of Theorem~\ref{main-theorem}.
It is also distinct from the classical limit, which effects a similar factorization in perturbation theory, cf. (\ref{classical}) below. 
The field renormalization is needed to ensure the non-triviality of (\ref{simple-example}), that is, to keep the $J$-dependent term in the exponent
non-zero and finite. Since $\fr{ \ti{C}_\N}{ \ti{C}_\N(0)^{1/2} }=Z_\N^{1/2} \fr{C_\N}{C_\N(0)^{1/2}}$, this can be achieved, e.g., by setting $Z_\N=C_\N(0)$.

Moving on to the proof of Proposition \ref{factorization-two}, we list the relevant properties of the covariance $C$. The  non-optimal bounds from Lemma \ref{propagator-lemma-universal} below are
convenient for us, as they facilitate uniform treatment of all dimensions $d\geq 2$. We postpone the proof to  Appendix \ref{properties-covariance}. 
\bel\label{propagator-lemma-universal} The propagator $C_\N$ has the following properties for $d\geq 2$:
\beqa
0<C_{\N}(x) \2<\2 C_\N(0)\quad \textrm{for} \quad x\neq 0, \label{bound-one}\\
C_\N(x) \2 \leq \2 \fr{c}{|x|^{d-3/2} }, \label{propagator-bound-zero}\\
  C_\N(0)\2\geq  \2 c_1\log(\N), \quad \textrm{for} \quad   {c_1>0}. 
 \label{propagator-bound}
\eeqa
\eel
Next, we introduce the following $\ell \times \ell$ matrices for  $\pmb{x}:=(x_1,\ldots, x_{\ell})\in \real^{\ell d}$, $0\leq \al\leq 1$,
 \beqa
\pmb{M}_{\al}(\pmb{x}):= I+\al \pmb{m}(\pmb{x}),
 \eeqa
where  $\pmb{m}_{i,j}(\pmb{x}):=\fr{C_\N(x_i-x_j)}{C_\N(0)}$ for $i\neq j$ and  $\pmb{m}_{i,i}(\pmb{x})=0$. We set $\pmb{M}(\pmb{x}):=\pmb{M}_{\al=1}(\pmb{x})$  and for $\ell=1$ it is understood that $\pmb{M}_{\al}(\pmb{x})=1$. Also, we will often abbreviate $ \pmb{M}_{\al}:=\pmb{M}_{\al}(\pmb{x})$, $\pmb{m}:=\pmb{m}(\pmb{x})$. 
\nin Next,  we define the following neighbourhood of the  diagonal in $\S^{\ell}$
\beqa
D_{\de}:=\{\, \pmb{x}\in \S^{\ell}  \,|\,  \exists_{i\neq j}  \textrm{ s.t. } |x_i-x_j|\leq \de  \,\}
\eeqa
and denote its complement in $\S^{\ell}$  by $D_{\de}'$.  (For $\ell=1$ we set $D_{\de}':= \S$).  We  choose $\de$, depending on $\N$, $\ell$, $d$, as follows
\beqa \label{de-def}
\de:=  \bigg(\fr{\ell}{ \sqrt{\log(\N)}} \bigg)^{\fr{1}{d-3/2}}.  \label{delta-def}
\eeqa
\newcommand{\ls}{s}
\bel\label{positive-definite-matrix}  For  $\pmb{x}\in  D_{\de}'$, and  $\N\geq \N_0$  (for some  $\N_0$ depending only on $d$, $m$)  the matrices $\pmb{M}_{\al}(\pmb{x})$ are {\redd positive definite}.  Furthermore, 
\beqa
\| \pmb{M}_{\al}(\pmb{x})^{-1}\|\leq c,\quad  \| \pmb{M}_{\al}(\pmb{x})^{-1}\pa_{\al} \pmb{M}_{\al}(\pmb{x})\| \leq  \fr{c}{\sqrt{\log(\N)}},  \label{M-bounds-x}
\eeqa
for $c$ possibly depending  on $d,m$ but  independent of $\N$, $\pmb{x}$.
\eel
\proof  For $\pmb{x}\in D_{\de}'$ we have  $|x_i-x_j|> \de$ for all $1\leq i< j\leq \ell $.  By Lemma~\ref{propagator-lemma-universal}, 
\beqa
\pmb{m}_{i,j}(\pmb{x})=\fr{C_\N(x_i-x_j)}{C_\N(0)}<    c \de^{-(d-3/2)} \fr{1}{\log(\N)} = \fr{c}{\ell} \fr{1}{\sqrt{\log(\N)}},
\eeqa
where we used definition~(\ref{de-def}).
Thus, recalling that $\pmb{m}_{i,i}=0$ and {\redd using Schur's test}
\beqa
\|\pmb{m}\|\leq  \sup_i\sum_{j}|\pmb{m}_{i,j}|\leq  c \fr{1}{\sqrt{\log(\N)}}\leq \h
 \label{matrix-element-bound}
\eeqa
for $\N\geq \N_0$, where $\N_0$ depends only on $d, m$.  Consequently, $\pmb{M}_{\al}=I+\al \pmb{m}$, $0\leq \al\leq 1$, is {\redd positive definite}. Furthermore, 
\beqa
\|\pmb{M}_{\al}^{-1}\|\2\leq\2 \sum_{\ell'=0}^{\infty} \| \pmb{m}\|^{\ell'},  \label{M-bound-one}\\
\|\pmb{M}_{\al}^{-1} \pa_{\al}\pmb{M}_{{\bluenew\al}}\|\2\leq \2 \|\pmb{M}_{\al}^{-1}\|\,\|\pmb{m}\|.  \label{M-bound-two}
\eeqa
Substituting (\ref{matrix-element-bound}) to (\ref{M-bound-one}), (\ref{M-bound-two}) we complete the proof. \qed
\bel\label{product-of-two} {\redd For $\pmb{x}\in D_{\de}'$ and $\La\geq \La_0$ as in Lemma~\ref{positive-definite-matrix},}  the following equality holds true 
\beqa
& & \fr{ \lan  \e^{\red \phi(J)} \thetas(\phi(x_1)) \ldots \thetas(\phi(x_{\ell}))\ran_{ \ti{C}_\N} }{\lan  \e^{\red \phi(J)}\ran_{\tiC_\N} }  \non\\
& &= \fr{1}{(2\pi)^{\ell/2}} \mrm{det}(  \pmb{M}(\pmb{x}) )^{-1/2}\int_{\real^{\ell}} d\pmb{w}\, \thetas(\tiC_\N(0)^{1/2}\pmb{w})   \e^{-\h  (\pmb{w} - \pmb{q})^T \pmb{M}(\pmb{x})^{-1}    (\pmb{w} -  \pmb{q})},
\eeqa
where we set $\thetas(\pmb{w}):=\thetas(w_1)\ldots \thetas(w_{\ell})$ and defined the vector
\beqa
\pmb{q}:=\bigg(  \fr{\lan  \de_{x_1}, \tiC_\N J\ran}{\tiC_\N(0)^{1/2}}, \ldots,  \fr{\lan \de_{x_\ell},\tiC_\N J\ran}{\tiC_\N(0)^{1/2}} \bigg).\label{q-def}
\eeqa
\eel
\proof Let us choose $\thetas_{\eps}\in S(\real)$ s.t.  $\thetas_{\eps} \nearrow \thetas$ pointwise. Then, using that (\ref{Fourier-transform-measure})
holds  for complex valued $J$ if we treat $\lan\,\cdot\,,\,\cdot\,\ran$ as complex-linear in both arguments,  
\beqa
\2 \2  \e^{\red -\h \lan J, \tiC_\N J\ran}\lan  \e^{\red \phi(J)} \thetas_{\eps}(\phi(x_1))\ldots \thetas_{\eps}(\phi(x_{\ell}))\ran_{ \ti{C}_\N}\non\\
\2=\2 \e^{\red -\h \lan J, \tiC_\N J\ran}\fr{1}{(2\pi)^{\ell/2}} \int_{\real^{\ell}} d{\pmb{u}}\, \wh{\thetas}_{\eps}(\pmb{u}) \lan \e^{\i \phi(\sum_{j=1}^{\ell} u_j\de_{x_j} {\red -\i J})} \ran_{\ti{C}_\N} \non\\
\2=\2 \e^{\red-\h \lan J, \tiC_\N J\ran}\fr{1}{(2\pi)^{\ell/2}} \int_{\real^{\ell}} d{\pmb{u}}\, \wh{\thetas}_{\eps}( \pmb{u})  \e^{-\h \lan \sum_{{\bluenew i}=1}^{\ell} u_{i}\de_{x_{i}}   {\red -\i J}, \ti{C}_\N  (\sum_{j=1}^{\ell} u_{j}\de_{x_{j}} {\red -\i J})\ran}  \non\\
\2=\2\fr{1}{(2\pi)^{\ell/2} } \tiC_{\N}(0)^{-\ell/2}\int_{\real^{\ell}}d\pmb{u} \, \wh{\thetas}_{\eps}(\tiC_\N(0)^{-1/2}\pmb{u} )  \e^{ -\h \sum_{j=1}^{\ell}u_j^2-\sum_{i<j} u_i u_j \fr{\tiC_\N(x_i-x_j)}{\tiC_\N(0)}+\i
\fr{ \lan \sum_{j=1}^{\ell}u_j\de_{x_j}, \ti{C}_\N J\ran }{\ti{C}_\N(0)^{1/2} }   }, \label{computations-below-x}
\eeqa
{\redd and we made a change of variables $\pmb{u} \mapsto  \tiC_\N(0)^{-1/2}\pmb{u}$ in the last step.}
Thus, we can rearrange this expression as follows, {\redd relying on the fact that $\pmb{M}(\pmb{x})$ is positive
definite for $\pmb{x}\in D_{\de}'$ by Lemma~\ref{positive-definite-matrix}}:
\beqa
\2 \2\fr{1}{(2\pi)^{ \ell/2} } \tiC_{\N}(0)^{-\ell/2}\int_{\real^{\ell}} d\pmb{u}\, \wh{\thetas}_{\eps}( \tiC_\N(0)^{-1/2}  \pmb{u} )  \e^{-\h \textbf{u}^T \pmb{M}(\pmb{x})\textbf{u}  
+\i\fr{ \lan \sum_{j=1}^{\ell}u_j\de_{x_j}  ,\tiC_\N J\ran}{\tiC_\N(0)^{1/2}}   }\non\\
\2 \2\phantom{4444444444}=\fr{1}{(2\pi)^{  \ell} } \tiC_{\N}(0)^{-\ell/2} \int_{\real^{\ell}} d\pmb{w}\, \thetas_{\eps}(\pmb{w}) \int_{\real^{\ell}} d\pmb{u}\,  \e^{-\i  \tiC_\N(0)^{-1/2} \pmb{w}^T \pmb{u} }  \e^{-\h \textbf{u}^T \pmb{M}(\pmb{x})\textbf{u}}
   \e^{\i \pmb{q}^T \pmb{u}   }\non\\
\2 \2\phantom{4444444444}=\fr{1}{(2\pi)^{  \ell} }  \tiC_{\N}(0)^{-\ell/2}\int_{\real^{\ell}} d\pmb{w}\, \thetas_{\eps}(\pmb{w})  \int_{\real^{\ell}} d \pmb{u}\,  \e^{-\i  (\tiC_\N(0)^{-1/2}\pmb{w} -  \pmb{q})^T   \pmb{u}}  \e^{-\h \textbf{u}^T \pmb{M}(\pmb{x})\textbf{u}} \non\\ 
\2  \2\phantom{4444444444}=\fr{1}{(2\pi)^{ \ell} } \int_{\real^{\ell}} d \pmb{w} \, \thetas_{\eps}(\tiC_\N(0)^{1/2} \pmb{w})   \int_{\real^{\ell}} d\pmb{u}\,  
\e^{-\i  (\pmb{w} -  \pmb{q})^T  \pmb{u}}  \e^{-\h \textbf{u}^T \pmb{M}(\pmb{x})\textbf{u}} \non\\ 
\2  \2\phantom{4444444444}= \fr{1}{(2\pi)^{\ell/2}} \mrm{det}(  \pmb{M}(\pmb{x}) )^{-1/2}\int_{\real^{\ell}} d\pmb{w}\, \thetas_{\eps}(\tiC_\N(0)^{1/2}\pmb{w}) \e^{-\h  (\pmb{w} -\pmb{q})^T \pmb{M}(\pmb{x})^{-1}    (\pmb{w} - \pmb{q}) }. \label{computations-below-one-x}
\eeqa
By taking the limit $\eps\to 0$ in the first line of (\ref{computations-below-x}) and in the last line of (\ref{computations-below-one-x}), and applying the dominated convergence 
we conclude the proof. \qed \\
\nin\textbf{Proof of Proposition~\ref{factorization-two}.}  First, we  estimate the volume of $D_{\de}$ for $\de$ given by (\ref{delta-def}).  We define $D_{\de;i,j}:=\{\, \pmb{x}\in \S^{\ell}\,|\, |x_i-x_j|\leq \de\,\}$. Clearly,
\beqa
D_{\de}\subset \bigcup_{i<j}D_{\de;i,j} \quad \Rightarrow \quad  |D_{\de}|\leq \sum_{i<j} |D_{\de;i,j}|. \label{sum-terms}
\eeqa
We note that  $|D_{\de;i,j}|\leq cL^{d(\ell-1)} \de^d$, hence $|D_{\de}|\leq c\ell^2 L^{d(\ell-1)} \de^d$.
Now we observe that
\beqa
\2 \2  \bigg| \int_{\S^{\ell}} d\pmb{x}\, \one_{D_{\de}}(\pmb{x}) \fr{  \lan {\red \e^{\phi(J)} }  \thetas(\phi(x_1)) \ldots \thetas(\phi(x_{\ell}))\ran_{ \ti{C}_\N} }{\lan \e^{\phi(J)} \ran_{\tiC_\N}  } \bigg| \non \\    
\2 \2\phantom{444444444444444444444444}\leq  c \|\F\|^{\ell}_{\infty} \ell^2 L^{d(\ell-1)}  \de^d 
\leq c   \|\F\|^{\ell}_{\infty}\ell^6 L^{d\ell}\fr{1}{\sqrt{\log(\N)}}, \label{diagonal-one}\\
\2 \2 \bigg|\int_{\S^{\ell}} d\pmb{x}\,  \one_{D_{\de}}(\pmb{x})  \fr{\lan   {\red \e^{ \phi(J)} }\thetas(\phi(x_1))\ran_{ \ti{C}_\N}}{\lan  
{\red \e^{\phi(J)} }  \ran_{ \ti{C}_\N}  } \ldots   \fr{ \lan   {\red \e^{\phi(J)} } \thetas(\phi(x_{\ell}))\ran_{ \ti{C}_\N} }{ \lan   {\red \e^{\phi(J)} } \ran_{ \ti{C}_\N}}\bigg|\non\\
\2  \2 \phantom{444444444444444444444444}\leq c \|\F\|^{\ell}_{\infty}   \ell^2 L^{d(\ell-1)}  \de^d \leq c\,  \|\F\|^{\ell}_{\infty}    \ell^6 L^{d\ell}\fr{1}{\sqrt{\log(\N)}},
 \label{diagonal-two}
 \eeqa
where we entered with the modulus under the Gaussian integrals and used  positivity of $\e^{\phi(J)}$. We also  exploited  that 
$\log(\La)\geq 1$, hence
\beqa
\de^d=\bigg( \fr{\ell}{\sqrt{\log(\N)}} \bigg)^{\fr{d}{d-3/2}} \leq \ell^{4}  \fr{1}{\sqrt{\log(\N)}}.
\eeqa
{\redd We  note that the bounds (\ref{diagonal-one}), (\ref{diagonal-two}) remain true if we replace $\one_{D_{\de}}(\,\cdot\,)$
under the integral by one and $\de^d$ on the r.h.s by $L^d$. Thereby we obtain the factorization property~(\ref{factorization-property}) 
for $\N\leq \N_0$, where $\N_0$ appeared in Lemma~\ref{positive-definite-matrix}. Thus, it suffices to consider $\N\geq \N_0 $  
in the following.}

Let us look at the $D_{\de}'$ part.  By Lemma~\ref{product-of-two}, the relevant quantity has the form:
\beqa
& &  \int_{\S^{\ell}} d\pmb{x}\,  \one_{D_{\de}'}(\pmb{x})\bigg[  \fr{ \lan {\red \e^{\phi(J)} } \thetas(\phi(x_1)) \ldots \thetas(\phi(x_{\ell}))\ran_{ \ti{C}_\N} }{\lan   {\red \e^{\phi(J)} } \ran_{ \ti{C}_\N} }-  
\fr{\lan   {\red \e^{\phi(J)} }\thetas(\phi(x_1))\ran_{ \ti{C}_\N}}{\lan  {\red \e^{\phi(J)} }  \ran_{ \ti{C}_\N}  } \ldots   \fr{ \lan   {\red \e^{\phi(J)} } \thetas(\phi(x_{\ell}))\ran_{ \ti{C}_\N} }{ \lan   {\red \e^{\phi(J)} } \ran_{ \ti{C}_\N}}  \bigg] \non\\
\2=\2 \int_{\S^{\ell}} d\pmb{x}\,  \one_{D_{\de}'}(\pmb{x})\bigg\{   (2\pi)^{-\ell/2} \det(\pmb{M})^{-1/2} 
\int_{\real^{\ell}} d\pmb{w}\, {\thetas}(\tiC_\N(0)^{1/2}\pmb{w})  \e^{-\h  (\pmb{w} -{\red  \pmb{q}}  )^T  \pmb{M}^{-1}   (\pmb{w}  -{\red  \pmb{q}} )   } \non\\
\2 \2 \phantom{444444444444444444444444444}-  (2\pi)^{-\ell/2} \int d\pmb{w}\, {\thetas}(\tiC_\N(0)^{1/2}\pmb{w})  \e^{-\h  (\pmb{w}  -{\red  \pmb{q}} )^T  (\pmb{w}  -{\red  \pmb{q}} )  }    \bigg\}. \label{off-diagonal}
\eeqa
We rewrite the expression in curly brackets as follows:   
\beqa
 \2 \2 (2\pi)^{-\ell/2}\int_0^1d\al \,\pa_{\al}\big[ \det(\pmb{M}_{\al} )^{-1/2} 
\int_{\real^{\ell}} d\pmb{w} \, {\thetas}(\tiC_\N(0)^{1/2}\pmb{w})  \e^{-\h  (\pmb{w}  -{\red   \pmb{q}}    )^T  \pmb{M}_{\al}^{-1} (\pmb{w}-{\red  \pmb{q}} )  } \big]\non\\
\2=\2 - \h(2\pi)^{-\ell/2}  \int_0^1d\al \,\big[   \det(\pmb{M}_{\al} )^{-1/2} \mrm{Tr}(\pmb{M}_{\al}^{-1}\pa_{\al} \pmb{M}_{\al})
\int_{\real^{\ell}} d\pmb{w}\, {\thetas}(\tiC_\N(0)^{1/2}\pmb{w}) \e^{-\h(\pmb{w}  -{\red   \pmb{q}}    )^T  \pmb{M}_{\al}^{-1} (\pmb{w}-{\red  \pmb{q}} )} \big]   \label{G-first} \\
\2 \2 +\h  (2\pi)^{-\ell/2} \int_0^1d\al \,\big[   \det(\pmb{M}_{\al} )^{-1/2} 
\int_{\real^{\ell}} d\pmb{w} \, {\thetas}(\tiC_\N(0)^{1/2}\pmb{w}) \times \non\\
\2 \2\phantom{44444444444444444444444444} \times   (\pmb{w}  -{\red   \pmb{q}})^T  \pmb{M}_{\al}^{-2}\pa_{\al}\pmb{M}_{\al}  (\pmb{w}  -{\red   \pmb{q}})   
\e^{-\h (\pmb{w}  -{\red   \pmb{q}}    )^T  \pmb{M}_{\al}^{-1} (\pmb{w}-{\red  \pmb{q}} )  } \big],  \label{G-second}
\eeqa
{\redd where we used that $\pmb{M}_{\al}(\pmb{x})$ is positive definite for $\pmb{x}\in D_{\de}' $}, thus  $\det(\pmb{M}_{\al} )^{-1/2}=\e^{-\h \mrm{Tr}(\log(\pmb{M}_{\al}))  }$. To estimate the respective terms above we bound $\F$ by its supremum and change
variables $\pmb{w}\mapsto \pmb{w}+\pmb{q} $
\beqa
|(\ref{G-first})| \2\leq\2 (2\pi)^{-\ell/2}\|\F\|^{\ell}_{\infty} \h \int_0^1d\al \,\big| \mrm{Tr}(\pmb{M}_{\al}^{-1}\pa_{\al} \pmb{M}_{\al}) 
\big| \int_{\real^\ell} d\pmb{w}\,   \e^{-\h  \pmb{w}^T   \pmb{w} }   \non\\
\2=\2 \h \|\F\|^{\ell}_{\infty} \int_0^1d\al \,\big| \mrm{Tr}(\pmb{M}_{\al}^{-1}\pa_{\al} \pmb{M}_{\al}) \big|  \leq 
 \h  \|\F\|^{\ell}_{\infty} \int_0^1d\al \,  \ell \|\pmb{M}_{\al}^{-1}\pa_{\al} \pmb{M}_{\al}\| \non\\
 \2\leq \2 c\ell \|\F\|^{\ell}_{\infty} \fr{1}{\sqrt{\log(\N)}},
  \eeqa
 where we used the  bound  (\ref{M-bounds-x}).
Similarly,
\beqa
|(\ref{G-second})| \2\leq\2 (2\pi)^{-\ell/2}\|\F\|^{\ell}_{\infty} \h \int_0^1d\al \,\| {\redd \pmb{M}_{\al}^{-1}} \pa_{\al} \pmb{M}_{\al}\| \int_{\real^{\ell}} d\pmb{w}\, \pmb{w}^T\pmb{w}\, \e^{-\h \pmb{w}^T\pmb{w}}   \non\\
\2=\2  \ell  \|\F\|^{\ell}_{\infty}  \h \int_0^1d\al \,\| {\redd \pmb{M}_{\al}^{-1}}\pa_{\al} \pmb{M}_{\al}\|  \leq c\ell {\bluenew  \|\F\|^{\ell}_{\infty}} \fr{1}{\sqrt{\log(\N)}}.
\eeqa
Thus, the claim follows from (\ref{diagonal-one})--(\ref{off-diagonal}).  \qed\\
\nin\textbf{Proof of Theorem~\ref{main-theorem}}.  We consider the functional $\Sm_\N(J)$ and expand  $ \e^{-\la \V(\phi)}$ into a power series in $\la$. Since the function 
$\thetas$ is bounded, the series is absolutely convergent for any $\la\in \real$.   We write
\beqa
\Sm_\N(J)\2=\2  \fr{ \lan \e^{\red \phi(J)}  \e^{-\la \V(\phi) } \ran_{\ti{C}_\N} }{ \lan  \e^{\red \phi(J)} \ran_{\ti{C}_\N}} 
=\exp\bigg(-\la \fr{\lan \e^{\red \phi(J)}  \V(\phi)\ran_{\ti{C}_\N} }{\lan \e^{\red \phi(J)}\ran_{\ti{C}_\N} } \bigg) \non\\
\2+\2\underbrace{  \sum_{\ell=0}^{\infty} \fr{(-\la)^{\ell}}{\ell!}  \bigg(   \fr{  \lan \e^{\red \phi(J)}   \V(\phi)^{\ell} \ran_{\ti{C}_\N} }{ \lan  \e^{\red \phi(J)} \ran_{\ti{C}_\N}} - 
   \bigg( \fr{\lan   {\blue \e^{\red \phi(J)} }\V(\phi) \ran_{ \ti{C}_\N}}{\lan  {\blue \e^{\red \phi(J)} }  \ran_{ \ti{C}_\N}  } \bigg)^{\ell}\bigg)}_{R_\N}. \label{rest-term}
\eeqa
By Proposition \ref{factorization-two}, we obtain that the rest term $R_\N$, defined by (\ref{rest-term}), satisfies
\beqa
|R_\N|\leq c  \fr{\e^{c_1 \la |\S|\, \|\F\|_{\infty} }}{ \sqrt{\log(\N)}}. \label{R-N-bound}
\eeqa
Consequently, 
\beqa
\Sm^{\mrm{c}}_\N(J)= \log(\Sm_\N(J)) ={\red -}\la \fr{\lan \e^{\red  \phi(J)}  \V(\phi)\ran_{\ti{C}_\N} }{\lan \e^{\red \phi(J)}\ran_{\ti{C}_\N} }{\red +}\log\bigg(1+\exp\bigg(\la \fr{\lan \e^{\red \phi(J)}  \V(\phi)\ran_{\ti{C}_\N} }{\lan \e^{\red \phi(J)}\ran_{\ti{C}_\N} } \bigg) R_\N\bigg). \label{final-formula-summand}
\eeqa
Now, making use of Lemma \ref{product-of-two} for $\ell=1$, we can write
\beqa
 \fr{\lan \e^{\red  \phi(J)}  \V(\phi)\ran_{\ti{C}_\N} }{\lan \e^{\red \phi(J)}\ran_{\ti{C}_\N} }= 
 \fr{1}{(2\pi)^{1/2}} \int_{\S} dx \int_{\real} dw\, \thetas((Z_\N C_\N(0))^{1/2}w )   \e^{-\h  \big(w - \big(\fr{Z_\N}{C_\N(0)}\big)^{1/2}  \lan  \de_{x}, C_\N J\ran \big)^2  }.
  \label{final-formula}
\eeqa
 This expression is bounded in $\N$, since we can estimate $\F$ by its supremum and then shift the $w$-variable.  
Thus,  the logarithm on the r.h.s. of  (\ref{final-formula-summand}) tends to zero  by estimate (\ref{R-N-bound}). 
{\bluenew Hence,} it suffices to compute the limit $\N\to \infty$ of (\ref{final-formula}) for the respective items of Theorem~\ref{main-theorem}:

\begin{enumerate}

\item[\ref{eta-large-negative}] For $-\infty<\eta<-1$,  $Z_\N=C_\N(0)^{\eta}$,   we have  $Z_\N/C_\N(0) \to 0$ and  $Z_\N C_\N(0)\to 0$. 
Then, by the dominated convergence and assumption~(\ref{limits-assumption-zero}),   
\beqa
(\ref{final-formula})\to\fr{1}{(2\pi)^{1/2}} \int_{\S} dx \int_{\real} dw\, [\F_+ \theta(w) + \F_-\theta(-w)]   \e^{-\h  w^2  }.
\eeqa
 As $\theta(w)=\h (\mrm{sgn}(w)+1)$, and $\mrm{sgn}$ is antisymmetric, we obtain the claim.

\item[\ref{eta-minus-one}] For $\eta=-1$ we have $Z_\N/C_\N(0) \to 0$  but $Z_\N C_\N(0) {\bluenew =} 1$. Then,  
\beqa
(\ref{final-formula})\2=\2 \fr{1}{(2\pi)^{1/2}} \int_{\S} dx \int_{\real} dw\, \thetas(w)   \e^{-\h  \big(w - \big(\fr{Z_\N}{C_\N(0)}\big)^{1/2}  \lan  \de_{x}, C_\N J\ran \big)^2  }\non\\
\2 \2\to \fr{1}{(2\pi)^{1/2}} \int_{\S} dx \int_{\real} dw\, \thetas(w)   \e^{-\h  w^2  }.
\eeqa

\item[\ref{eta-middle}] For $-1<\eta<1$ we have $Z_\N/C_\N(0) \to 0$  and  $Z_\N C_\N(0)\to \infty$.  Then,  by 
assumption~(\ref{limits-assumption}),
\beqa
(\ref{final-formula})\to \fr{1}{(2\pi)^{1/2}} \int_{\S} dx \int_{\real} dw\, [\F^+ \theta(w) + \F^-\theta(-w)]   \e^{-\h  w^2  }.
\eeqa

\item[\ref{eta-one}] For $\eta=1$ we have  $Z_\N/C_\N(0) {\bluenew =} 1$ and  $Z_\N C_\N(0)\to \infty$. Hence, again by (\ref{limits-assumption}),
\beqa
(\ref{final-formula}) \to \fr{1}{(2\pi)^{1/2}} \int_{\S} dx \int_{\real} dw\,[\F^+ \theta(w) + \F^-\theta(-w)] \e^{-\h  (w -   \lan  \de_{x}, C_\N J\ran )^2  }.
\eeqa

\item[\ref{eta-large-positive}] For $1<\eta<\infty$ we have   $Z_\N/C_\N(0) \to \infty$ and $Z_\N C_\N(0)\to \infty$. Then, assuming that $\hat{J}$ is compactly
supported, vanishing of  $x\mapsto \lan \de_x, CJ\ran $ on a set of non-zero Lebesgue measure implies $J=0$. Thus, inserting 
$1=\one( \lan \de_x, CJ\ran=0)+\one( \lan \de_x, CJ\ran \neq 0)$ and using (\ref{limits-assumption}), we obtain
\beqa
(\ref{final-formula})\2=\2\fr{1}{(2\pi)^{1/2}} \int_{\S} dx \int_{\real} dw\, \thetas((Z_\N C_\N(0))^{1/2}(w + \bigg(\fr{Z_\N}{C_\N(0)}\bigg)^{1/2}  \lan  \de_{x}, C_\N J\ran ))   \e^{-\h  w^2  }\non\\
\2\to\2 \one(J=0)\fr{1}{(2\pi)^{1/2}} \int_{\S} dx\,  \int_{\real} dw\, (\F^+ \theta(w) + \F^-\theta(-w)  )   \e^{-\h  w^2  } \non\\
\2 \2 + \one(J\neq 0)\int_{\S} dx\,  \big(\F^+ \theta(  \lan  \de_{x}, CJ\ran) + \F^-\theta(-  \lan  \de_{x}, CJ\ran)  \big). 
\eeqa
After obvious rearrangements this gives the claim. (Without the assumption that $\hat{J}$ is compactly supported, 
the limit in the $\lan \de_x, CJ\ran=0$ case would depend on the convergence properties of the sequences $\N\mapsto \big(\fr{Z_\N}{C_\N(0)}\big)^{1/2}  \lan  \de_{x}, (C_\N-C)J\ran $. These
may a priori vary in a complicated manner with $x$ and $J$).
\qed

\end{enumerate}
One unfortunate  feature of Theorem~\ref{main-theorem} 
is the absence of any $J$ dependence of $\hat{S}^{\mrm{c}}(J)$
 in parts  \ref{eta-large-negative}--\ref{eta-middle}, i.e., for $\eta<1$.
With the {\blues (massive)}  sine-Gordon model in mind, one could hope that such dependence emerges if we relax the assumptions
(\ref{limits-assumption-zero}), (\ref{limits-assumption}),  e.g., by  admitting some oscillations of $\F$ near
zero or infinity.   It turns out that  this would not help. In fact,
let us set $t_\N:=C_\N(0)^{1/2}$, $Z_\N=C_\N(0)^{\eta}=t_\N^{2\eta}$ and $a(x):=\lan  \de_{x}, C_\N J\ran$. Then, by (\ref{final-formula}),
we can write
\beqa
\la \fr{\lan \e^{\red  \phi(J)}  \V(\phi)\ran_{\ti{C}_\N} }{\lan \e^{\red \phi(J)}\ran_{\ti{C}_\N} }= 
 \fr{\la}{(2\pi)^{1/2}} \int_{\S} dx \int_{\real} dw\, \thetas( t_\N^{\eta+1} w )   \e^{-\h  (w -  t_\N^{\eta-1} a(x) )^2  }.
  \label{final-formula-x}
\eeqa
{\redd Now we introduce a parameter $s\in [0,1]$, replace $a(x)$ with $sa(x)$ on the r.h.s. of (\ref{final-formula-x}) and call the
resulting function $f_{\N}(s)$. By estimating $|f_{\N}(1)-f_{\N}(0)|\leq \int_0^1 ds |\pa_s f_{\N}(s)|$ and taking the
limit $\N\to \infty$, we obtain that the $J$ dependence is trivial for $\eta<1$ relying only on the boundedness of $V$. This follows from} 
\beqa
\pa_s\int_{\real} dw\, \thetas( t_\N^{\eta+1} w )   \e^{-\h  (w -  t_\N^{\eta-1} s a(x) )^2  }= t_\N^{\eta-1} a(x) \int_{\real} dw\, \thetas( t_\N^{\eta+1} w )  \pa_w \e^{-\h  (w -  t_\N^{\eta-1} s a(x) )^2} \label{obstacle}
\eeqa
due to the pre-factor $t_\N^{\eta-1}$. One could try to compensate this pre-factor by choosing $\la$ growing with $\N$, but such a growth would immediately spoil the bound (\ref{R-N-bound}) and thus undermine the factorization mechanism from Proposition~\ref{factorization-two}.
Incidentally, this is how the sine-Gordon model $\F(\phi)=\!-\!:\!\cos(\be \phi)\!:$ evades our analysis: The Wick ordering gives {\bluenew the} effective behaviour of the coupling constant $\la\sim \e^{\h \be C_\N(0)}$ which is incompatibe with the factorization.  


\section{Generalization} \label{generalisation}
\setcounter{equation}{0}
\newcommand{\thet}{\kappa}

It is clear from estimate (\ref{R-N-bound}) that our methods apply 
to interaction functions $\F$ depending on $\N$, as long as their supremum norms remain uniformly bounded in $\N$.
We consider {\bluenew a} $\N$ dependent interaction function of the form $\F_{\N}(w):=\F(z_{\N} w)$
and the corresponding interaction term $\V_{\N}(\phi):=\int_B dx\, \F_{\N}(\phi(x))$. 
We can compute the UV limit of the resulting modified generating functional, which we still denote by $\Sm^{\mrm{c}}_{\N}(J)$,
using obvious modifications of formulas (\ref{final-formula-summand}), (\ref{final-formula}).
Thus, the problem reduces to computing the limit $\N\to \infty$ of
\beqa
 \fr{\lan \e^{\red  \phi(J)}  \V_{\N}(\phi)\ran_{\ti{C}_\N} }{\lan \e^{\red \phi(J)}\ran_{\ti{C}_\N} }= 
 \fr{1}{(2\pi)^{1/2}} \int_{\S} dx \int_{\real} dw\, \thetas(z_{\N}(Z_\N C_\N(0))^{1/2}w )   \e^{-\h  \big(w - \big(\fr{Z_\N}{C_\N(0)}\big)^{1/2}  \lan  \de_{x}, C_\N J\ran \big)^2}.
  \label{final-formula-gen}
\eeqa
Setting for concreteness $z_{\La}:=C_{\N}(0)^{\thet}$,  $\thet\in \real$,  Theorem~\ref{main-theorem}
can be generalized\footnote{Strictly speaking, in part (e) additional regularity assumptions on $V$ may be needed
to control the entire range of $\thet$.}. We focus here on part (d), as it brings interesting new insights. It splits into three cases depending on $\ka$:
\bec\label{main-theorem-gen} Set $Z_\N=C_\N(0)$, $z_{\N}=C_{\N}(0)^{\thet}$, $\thet\in \real$. Then, the limit $\Sm^{\mrm{c}}(J):=\lim_{\N\to \infty}\Sm^{\mrm{c}}_\N(J)$ exists for any $J\in S(\real^d;\real)$  under the assumptions  specified below. It has the form:
\begin{enumerate}[label={(\alph*)},itemindent=1em]

\item[(d.1)] Let $\thet<-1$. Then, assuming (\ref{limits-assumption-zero}),
\beqa
\Sm^{\mrm{c}}(J)={\red-} \la |\S|\fr{\F_++\F_-}{2}  {\red -}\la \fr{\F_+-\F_-}{2}   \fr{1}{(2\pi)^{1/2}}\int_{\S} dx\,\int dw\, \mrm{sgn}(w) \e^{-\h (w- \lan \de_x,  C  J\ran )^2}.
\eeqa

\item[(d.2)] Let $\thet=-1$. Then, without additional assumptions,
\beqa
\Sm^{\mrm{c}}(J)= {\red -}\la   \fr{1}{(2\pi)^{1/2}}\int_{\S} dx\,\int dw\, \F(w) \e^{-\h (w- \lan \de_x,  C  J\ran )^2}.
\eeqa

\item[(d.3)] Let $\thet>-1$. Then, assuming (\ref{limits-assumption}),
\beqa
\Sm^{\mrm{c}}(J)={\red-} \la |\S|\fr{\F^++\F^-}{2}  {\red -}\la \fr{\F^+-\F^-}{2}   \fr{1}{(2\pi)^{1/2}}\int_{\S} dx\,\int dw\, \mrm{sgn}(w) \e^{-\h (w- \lan \de_x,  C  J\ran )^2}.
\eeqa

\end{enumerate}

\eec
We note that in the case \emph{(d.1)} we obtain Schwinger functions of the form  (\ref{Schwinger-functions})
up to the coupling constant renormalization, which is now $\fr{\F_+-\F_-}{2}$. Hence,  non-Gaussianity holds
if and only if $\F$ is discontinuous at zero. Thus, our theories  turn out  to be sensitive to local regularity properties of
the interaction functions. 

In the case \emph{(d.2)} the Schwinger functions differ more substantially from  (\ref{Schwinger-functions}). By repeating
the steps from Appendix~\ref{Schwinger-functions-app}, we obtain, for $n\neq 2$,
\beqa
S^{\mrm{c}}_{n}(x_1,\ldots, x_n)
\2=\2{\red -}\la \bigg( \fr{1}{(2\pi)^{1/2} }\int dw\, (\pa^nV)(w)\e^{-\h w^2}\bigg) \int_{\S}dx \,   C(x-x_1)\ldots C(x-x_n).
\label{Schwinger-functions-gen}
\eeqa
These $n$-point  functions coincide with the tree level one-particle irreducible Schwinger functions of a perturbative
theory with the interaction function given by the convolution $V\!*\!\fr{1}{\sqrt{2\pi}} \e^{-\h (\,\cdot\,)^2}$.  Thus, the class
of theories tractable by our methods is richer than one could expect from (\ref{Schwinger-functions}).

\section{Conclusion and outlook}\label{conclusion}
\setcounter{equation}{0}

In this paper we succeeded in computing exactly the  UV limits $\Sm^{\mrm{c}}(J)$ of the modified generating 
functionals~(\ref{generating-functionals}) for a large class of bounded measurable interaction functions $\F$. 
We computed the connected Schwinger functions $S^{\mrm{c}}_{n\neq 2}$ as derivatives of $\Sm^{\mrm{c}}(J)$ w.r.t. $J$ and observed that they coincide with the tree-level one-particle irreducible Schwinger functions of the $\mrm{erf}(\phi/\sqrt{2})$ theory 
with the coupling constant renormalized by the factor $\fr{V^+-V^{-}}{2}$. {\redd In Section~\ref{generalisation} we noted that $\F$ may depend on $\N$ as long as it is uniformly bounded in the supremum norm.  In the class of interactions $\phi \mapsto \F(z_\N\phi)$,
we could change the coupling constant renormalization to $\fr{V_+-V_{-}}{2}$ so that the Schwinger functions probe the continuity of
$\F$ at zero}.

A more intriguing direction is to look at functions $\F$ which become increasingly rough as $\N$ increases, but remain uniformly bounded in $\La$. For
example, the number of discontinuities could increase with $\N$.  The hope is that thinking in this direction one could
invalidate    (\ref{obstacle})  by  disintegration of the $w$-integrals and obtain non-trivial Schwinger functions $S^{\mrm{c}}_{n\neq 2}$ also for $\eta<1$. This would be an important step forward, as only for $\eta\leq 0$ a finite  two-point
connected Schwinger function is immediately available, as we now explain.

Strictly speaking, the limiting modified functionals $\Sm^{\mrm{c}}(J)$ do not carry any information about the two-point
function of the limiting theory. But it appears consistent with (\ref{generating-functionals}), (\ref{Fourier-transform-measure})  to define the two-point connected Schwinger function by
\beqa
S^{\mrm{c}}_{2}(x_1, x_2):=\fr{\de}{\de J(x_1)}  \fr{\de}{\de J(x_2)}\Sm^{\mrm{c}}(J) |_{J=0}+ \lim_{\N\to \infty}Z_\N C_\N(x_1,x_2). \label{two-point-connected}
\eeqa
{\bluenew Since $Z_\N=C_{\N}(0)^{\eta}$}, expression~(\ref{two-point-connected}) is finite and well defined only for ${\redd \eta\leq 0}$, in which case {\bluenew all the other connected Schwinger functions vanish, i.e.,} $S^{\mrm{c}}_{n\neq 2}=0$. 
For ${\redd \eta<0}$ we have $S^{\mrm{c}}_{2}(x_1, x_2)=0$,
i.e., all the Schwinger functions are zero, whereas for {\redd $\eta=0$} we obtain $S^{\mrm{c}}_{2}(x_1, x_2)=C(x_1,x_2)$ and reproduce the free field theory.  In the case of $\eta=1$, in which we obtained the Schwinger functions~(\ref{Schwinger-functions}), we have $S^{\mrm{c}}_{2}(x_1, x_2)=\infty$. We observe that this divergence coexists with relation (\ref{UV-stability}), which demonstrates the limitations of the conventional UV stability criterion.

From the point of view of Euclidean constructive QFT the blow-up of the two-point function 
is an obstacle, which undermines the Osterwalder-Schrader reconstruction of the model as a Minkowskian QFT. From the Minkowskian perspective
 this divergence may be a manifestation of the fact that  the vacuum vector is not in the domain of the quantum field or
that pointlike-localized fields do not exist at all in these models. Actually, there are general arguments pointing in a similar direction: if $Z_\N>1$ or $d> 4$ then 
interacting scalar fields must violate the canonical commutation relations \cite[Eq (10.7.22)]{We},\cite{Ba87}.
As a singular behaviour of pointlike localized  fields is not an obstacle to the existence of a QFT in
the Haag-Kastler sense, it may be fruitful  to pursue a direct Minkowskian construction of models with bounded interactions
 in the spirit of \cite{BF20, BFR21}. 

Nevertheless, it would be preferable to tame the blow-up of the two-point function. As the problem 
occurs at the level of  free theory, it may be solvable by adding some spectator field, not interacting with the original system. To
 illustrate the idea, we note that the denominator  in (\ref{generating-functionals}) satisfies $\fr{1}{S_{0,\N}(J)}=\lan \e^{\i \vp(J)} \ran_{\tiC_\N}$, {\redd thus differs from $S_{0,\N}(J)$ only by the imaginary unit in the exponent.}
Thus, we can express  $\Sm_\N(J)$ as a  (non-modified) generating functional of the complex field $\psi:=\phi+\i\vp$ and interpret the $n$-point functions~(\ref{Schwinger-functions}), including $n=2$, as connected Schwinger functions of $\psi$. 
By an analogous construction we obtain the Schwinger functions of $\psi^*$, but $n$-point functions involving
both $\psi$ and $\psi^*$ are not immediately available. Their construction amounts to the problem of coupling of
two probability measures, whose solution is, in general, not unique. While naive choices lead to the blow-up 
of such Schwinger functions, one can hope for a solution consistent with Osterwalder-Schrader axioms.

Another direction is to take simultaneously the UV limit and the classical limit, i.e., to  analyse
\beqa
\check{S}^{\mrm{c}}_{\N}(J):= \hbar_\N \log \lan \e^{ \hbar_\N^{-1}\phi(J)} \e^{-\la \hbar_\N^{-1} \V(\phi) } \ran_{ {\redd \hbar_\N} C_\N}. \label{classical}
\eeqa
It is easy to show by our methods that  $\check{S}^{\mrm{c}}_{\N}(J)$ is bounded in $\N$ for  $\hbar_\N=C_\N(0)^{-1}$. 
Based on preliminary computations, we conjecture that  it has a limit and $\lim_{\N\to\infty}\check{S}^{\mrm{c}}_{\N}(J)=\Sm^{\mrm{c}}(J) +\h \lan J, CJ\ran$. In this case all the
Schwinger functions are finite and the family (\ref{Schwinger-functions}) is naturally complemented by the free covariance $C$. This is 
reminiscent of formal perturbation theory, where the classical limit extracts the tree-level connected Schwinger functions.
Unfortunately, the Osterwalder-Schrader reconstruction {\redd would  fail to give a non-trivial QFT}, due to the Jost-Schroer theorem.
But the construction would {\bluenew provide} an interesting, non-Gaussian, statistical physics model.  


\appendix

\section{Existence of the functional measure with cut-offs}\label{Path-integral}
\setcounter{equation}{0}

We define the Sobolev spaces $L^{2,s}(\real^d):=\{\, \phi\in S'(\real^d)\,|\, (1-\De)^{s/2}\phi \in L^2(\real^d)\,\}$ for $s\in \nat$. 
\bel\label{concentration-lemma} Let $\chi\in C_0^{\infty}(\real^d)$ be an approximate characteristic function of the set $\S$.
Then, the Gaussian measure $\nu_{\tiC_\N}$  is concentrated on $\phi$ s.t.
$\chi\phi\in L^{2,s}(\real^d)$ for any $s\in \nat$.
\eel
\proof  For any $c>0$
\beqa
\int d\nu_{ \ti{C}_\N }(\phi)\, \one(  \|  (1-\De)^{s/2} \chi \phi\|^2_2 =\infty   )  \2\leq\2 \fr{1}{c} \int d\nu_{\ti{C}_\N}(\phi) \, \| (1-\De)^{s/2}\chi\phi\|^2_2\non\\
\2=\2 \fr{1}{c} \int dx  \,\chi(x) \int d\nu_{ \ti{C}_\N }  (\phi) \,  (1-\Delta_{x'})^s \chi(x')\phi(x) \phi(x')|_{x'=x}\non\\
\2=\2  \fr{1}{c} \int dx \, \chi(x)\,  (1-\Delta_{x'})^s \chi(x')\ti C_\N(x-x')|_{x'=x}. \label{concentration}
\eeqa
Recalling formula (\ref{covariance}) for $C_\N$, we obtain that (\ref{concentration}) is finite for any fixed $\N$.
As $c$ can be chosen arbitrarily large, 
this completes the proof. \qed\\
We recall the Sobolev embedding theorem \cite[Theorem 1.66]{BCD}:
\bet\label{Sobolev-embedding} The space $L^{2,s}(\real^d)$ embeds continuously in  the
H\"older space $C^{k,\rho}(\real^d)$ if  $s\geq d/2+k+\rho$ for some
$k\in \nat$ and $\rho\in ]0,1[$.  
\eet
\nin For the H\"older space $C^{k,\rho}(\real^d)$ we refer to \cite[Definition 1.49]{BCD}. We only need here
that it consists of continuous functions.

\bec For any bounded measurable function $\F$ the {\bluenew probability} measure 
\beqa
d\mu_{\tiC_\N}(\phi)=\fr{1}{\mathcal{Z}_\N}\exp\big(-\la \int_{\S} d x\, \thetas(\phi(x)) \big) d\nu_{\tiC_\N}(\phi),
\eeqa
where $\mathcal{Z}_\N$ is the normalization constant, is well defined.
\eec
\proof We note that $\int_{\S} dx\, \thetas(\phi(x))= \int_{\S} dx\, \thetas((\chi\phi)(x))$.
By Lemma~\ref{concentration-lemma}, we can assume that $\chi\phi\in L^{2,s}(\real^d)$.   
 Thus, by Theorem~\ref{Sobolev-embedding}, it has a continuous representative 
which we can compose with $\thetas$. By the Jensen inequality, boundedness of $\F$ and compactness of $\S$ we have $\mathcal{Z}_\N>0$. \qed
\section{Computation of Schwinger functions} \label{Schwinger-functions-app}
\setcounter{equation}{0}

Recall the generating functional (\ref{Generating-functional-thm})
\beqa
\Sm^{\mrm{c}}(J)={\red-} \la |\S| \fr{\F^++\F^-}{2} {\red -}\la \fr{\F^+-\F^-}{2}   \fr{1}{(2\pi)^{1/2}}\int_{\S} dx\,\int dw\, \mrm{sgn}(w) \e^{-\h (w- \lan \de_x,  C  J\ran )^2}.
\eeqa
We can write, by the Plancherel theorem,
\beqa
\int dw\, \mrm{sgn}(w) \e^{-\h (w- \lan \de_x,  C  J\ran )^2}=\int du\, \wh{\mrm{sgn}}(u) \e^{-\h u^2 } \e^{\i u \lan \de_x,  C  J\ran},
\eeqa
which is the action of the distribution $\wh{\mrm{sgn}}$ on a test function. As we are only interested in Schwinger functions
smeared with test functions $f_1,\ldots, f_n\in S(\real^d;\real)$, it suffices to compute functional derivatives w.r.t. $J$ at zero in the direction of such functions.
This amounts to:
\beqa
\pa_{\eps_1}\ldots \pa_{\eps_n} \int du\, \wh{\mrm{sgn}}(u) \e^{-\h u^2 } \e^{\i u \sum_{i=1}^n\eps_i\lan \de_x,  C  f_i\ran}|_{\eps_1,\ldots, \eps_n=0}.
\eeqa
It is easily seen that these derivatives commute with the action of $\wh{\mrm{sgn}}$ as the approximating sequences converge in $S(\real^d;\real)$.
This expression equals
\beqa
 \int du\, (\i u)^n\wh{\mrm{sgn}}(u) \e^{-\h u^2 }  \lan \de_x,  C  f_1\ran \ldots  \lan \de_x,  C  f_n\ran.
\eeqa
We note the simple identities
\beqa
 \int du\, (\i u)^n\wh{\mrm{sgn}}(u) \e^{-\h u^2 } \2=\2 2\int du\,  \wh{(\pa^{n-1}  \de)}(u) \e^{-\h u^2 } \non\\
\2=\2 2\int dw\,  (\pa^{n-1}  \de)(w) \e^{-\h w^2 }\non\\
\2=\2 2(-1)^{n-1} (\pa^{n-1} \e^{-\h w^2 })|_{w=0}=\sqrt{2\pi}  (\pa^n_w \mrm{erf}(w/\sqrt{2}))|_{w=0},
 \eeqa
where we used the definition of the error function   $\mrm{erf}(w):=\fr{2}{\sqrt{\pi}} \int_0^w dw'\, \e^{-(w')^2}$ 
and the fact that it is antisymmetric.

\section{Proof of Lemma~\ref{propagator-lemma-universal}} \label{properties-covariance}
\setcounter{equation}{0}

By performing the $p_0$-integral in (\ref{covariance}) we obtain
\beqa
C_\N(x)=\int \fr{d \vec{p}}{(2\pi)^d} \int_{1/\N^2}^{\infty} d\al\, \sqrt{\fr{\pi}{\al} }\e^{-\al( (\vec{p})^2+m^2 )  }  \e^{-\h (2\al)^{-1} |x|^2}, \label{C-N-rep-xx}
\eeqa
 hence $0<C_\N(x)$ and $C_\N(x)<C_\N(0)$ for $x\neq 0$.  Property (\ref{propagator-bound-zero}) follows directly from \cite[Prop. 7.2.1]{GJ}.
 To obtain (\ref{propagator-bound}), we compute also the $\vec{p}$-integral in (\ref{C-N-rep-xx}):
 \beqa
 C_\N(x) = \fr{1}{2^d\pi^{d/2}} \int_{1/\N^2}^{\infty} d\al\,     \al^{-d/2} \e^{-\al m^2   }  \e^{-\h (2\al)^{-1} |x|^2}. 
 \eeqa
Consequently   $C_\N(0)\geq  c \int_{1/\N^2}^{2} d\al\,     \al^{-d/2}$ for some $c>0$.


\end{document}